\def \be {\begin{equation}}
\def \ee {\end{equation}}
\def \bea {\begin{eqnarray}}
\def \eea {\end{eqnarray}}
\def \nn {\nonumber}

\def \Z {{\bf Z}}
\def \del {\partial}
\def \dels {\partial\kern-.5em / \kern.5em}
\def \As {{A\kern-.5em / \kern.5em}}
\def \Ds {D\kern-.7em / \kern.5em}

\def \a {\alpha}

\def \dag {\dagger}

\def \d {\delta}
\def \eps {\epsilon}

\def \lam {\lambda}
\def \Lam {\Lambda}
\def \s {\sigma}

\def \om {\omega}

\def \t {\tau}

\documentclass[12pt]{article}

\begin{document}
\begin{titlepage}
%\catcode`\@=11
%\catcode`\@=12
%\twocolumn[\hsize\textwidth\columnwidth\hsize\csname%
%@twocolumnfalse\endcsname

%\draft
\begin{center}
\hfill hep-th/0208182\\
\vskip .5in

\textbf{\large
Virasoro Algebra for Particles \\
with Higher Derivative Interactions
}

\vskip .5in
{\large Pei-Ming Ho} %$^1$}
\vskip 15pt

{\small Department of Physics,
National Taiwan University,
Taipei, Taiwan, R.O.C.}\\
% {\small \em $^2$Department of Physics,
% Tamkang University, Tamsui, Taiwan, R.O.C.}

\vskip .2in
\sffamily{
pmho@phys.ntu.edu.tw}

\vspace{60pt}
%\maketitle
\end{center}
\begin{abstract}

In this paper we show that
the worldline reparametrization for particles
with higher derivative interactions appears as
a higher dimensional symmetry,
which is generated by the truncated Virasoro algebra.
We also argue that for generic nonlocal particle theories
the fields on the worldline may be promoted to
those living on a two dimensional worldsheet,
and the reparametrization symmetry becomes
locally the same as the conformal symmetry.

\end{abstract}
%\pacs{PACS numbers: 11.25.-w, 11.25.Mj, 11.25.Sq}%]
\end{titlepage}
%\begin{narrowtext}
\setcounter{footnote}{0}

\section{Introduction}

Due to technical difficulties,
physicists have been reluctant to
consider higher derivative interactions.
Yet they are unavoidable in many important physical problems.
For a partial list, see \cite{CHY}.
In string theory, for example,
we do not fully understand how to deal with
higher derivative terms in the worldsheet action
as they are nonrenormalizable,
even though such background interactions should exist.
String field theory is another outstanding example
of nonlocal theory.
For some recent interests in nonlocal theories
in the context of string theory, see \cite{string}.

In fact, even for particles,
higher derivative interactions are not well understood.
In this paper we study a very basic property of
particle worldline theory --
the reparametrization symmetry,
for worldline Largrangians with higher derivatives.
We find that, remarkably,
while the phase space of higher derivative theories
are of higher dimensions,
the reparametrization group acting on the phase space
is also of higher dimension.
Furthermore,
for generic nonlocal theories,
the reparametrization group is locally equivalent
to the conformal group of two dimensions.
It will be very interesting to see whether
a new class of well-defined conformal field theories
can originate from nonlocal worldline theories.

The plan of this paper is as follows.
We first review the worldline theory of a charged particle
and define its higher derivative generalization
(sec.\ref{Review}).
Then we show that both the phase space and
the reparametrization group
are of higher dimensions for higher derivative theories
(sec.\ref{Symmetry}).
In the nonlocal limit,
the reparametrization group becomes
the conformal group (sec.\ref{Conformal}).
%After giving an example (sec.\ref{Example}),
Finally we comment on the connection to the string theory
and the generalization of general covariance,
as well as on the issue of stability
(sec.\ref{Remarks}).

\section{Worldline Lagrangian}
\label{Review}

Let us start by reviewing the worldline theory
of a charged particle moving in curved spacetime
with metric $g_{\mu\nu}$
\be \label{S0}
S_0 = \frac{1}{2} \int d\tau
( e^{-1} g_{\mu\nu}\dot{x}^{\mu}\dot{x}^{\nu}
+ A_{\mu}(x) \dot{x}^{\mu}
- e \phi(x) ),
\ee
where $e = e(\tau)$ is the tetrad of the worldline
and $\phi$ is the potential term.
For particles of mass $m$ without external potential,
$\phi = m^2$.
The action is invariant under the reparametrization of
the worldline parameter $\tau$
\be \label{reparatau}
\tau \rightarrow \tau' = \tau-\eps(\tau).
\ee
It induces a transformation of the worldline fields $e$ and $x$
\be
\d e = \frac{d}{d\t}(\eps e), \quad \mbox{and} \quad
\d x = \eps \dot{x}.
\label{repara}
\ee

The symmetry of general coordinate transformation
in the gravitational theory is reflected in the particle theory
as the invariance of the action under
a field redefinition
$ x^{\mu} \rightarrow x'{}^{\mu}(x) $
and a simultaneous transformation of the coefficients
$g_{\mu\nu}$ and $A_{\mu}$.
%\be
%g_{\mu\nu} \rightarrow
%g'_{\mu\nu} = \frac{\del x^{\kappa}}{\del x'{}^{\mu}}
%\frac{\del x^{\lam}}{\del x'{}^{\nu}} g_{\kappa\lam}, \quad
%A_{\mu} \rightarrow
%A'_{\mu} = \frac{\del x^{\nu}}{\del x'{}^{\mu}} A_{\nu}.
%\ee
The $U(1)$ symmetry of the gauge potential $A_{\mu}$,
on the other hand,
appears as changing the Lagrangian by a total derivative.

The only generator of the reparametrization symmetry
is the Hamiltonian $H$.
In the usual covariant quantization scheme,
we fix the gauge by setting $e = 1$
and impose the constraint $ H = 0 $.

Let us now consider the worldline Lagrangian
in its full generality.
The most general particle Lagrangian
$L(x, \dot{x}, \ddot{x}, \cdots; e)$
is an arbitrary function of $x^{\mu}$
and all of their time derivatives.
We can expand $L$ by the number of time derivatives as
\bea
L &=& e \left[
A^{(0)}(x) + A^{(1)}_{\mu}(x) D x^{\mu}
+ ( A^{(01)}_{\mu}(x) D^2 x^{\mu}
+ \frac{1}{2}A^{(20)}_{\mu\nu}(x) D x^{\mu} D x^{\nu} )
\right. \nn \\
&& \left.
+ ( A^{(001)}_{\mu} D^3 x^{\mu}
+ A^{(110)}_{\mu\nu} D^2 x^{\mu} D x^{\nu}
+ A^{(300)}_{\mu\nu\lam} D x^{\mu} D x^{\nu} D x^{\lam} )
+ \cdots
\right],
\label{totalL}
\eea
where
\be \label{D}
D \equiv e^{-1}\frac{d}{d\tau}
\ee
is the covariant derivative of $\tau$.
For a given order $n$ of $D$
and a set of integers
$P(n)=\{ P_k\geq 0; k=1,\cdots,n \}$
satisfying
$ \sum_{k=1}^{n} k P_k = n $,
there is a spacetime field $A^{(P_1,\cdots,P_n)}(x)$ coupled
to the particle by the interaction
\be
L_{P(n)} \equiv e A^{P(n)}
= e \left( A^{P(n)}_{\mu_1\cdots\mu_m} \;
\Pi_{i=1}^{P_1} D x^{\mu_i} \;
\Pi_{j=1}^{P_2} D^2 x^{\mu_{P_1+j}} \;
\Pi_{k=1}^{P_3} D^3 x^{\mu_{P_1+P_2+k}}
\cdots
\right)
\ee
invariant under (\ref{repara}),
where $m=\sum_{k=1}^n P_k$.

\section{Reparametrization Symmetry}
\label{Symmetry}

We start with Lagrangians (\ref{totalL})
involving only finite order time derivatives.
Assume that the Lagrangian (\ref{totalL}) is
a function of $\{x^{(n)};\;\; n = 0,1,\cdot,N\}$,
where we used the notation
$ f^{(n)} \equiv \left(\frac{d}{d\tau}\right)^n f(\tau) $,
and that it is nondegenerate.
To apply canonical quantization,
we can introduce new variables $x_1,\cdots,x_{N-1}$ and add
\be \label{L1}
L' = \sum_{i=1}^{N-1} \lam_i (x_i-\dot{x}_{i-1})
\ee
to the Lagrangian,
where $\lam_i$'s are the Lagrange multipliers.
This trick allows us to replace $x^{(n)}$ by $x_n$
for $n = 1,2,\cdots,N-1$ in the Lagrangian,
which then becomes a function of
$x_0, x_1, \cdots, x_{N-1}$ and $\dot{x}_{N-1}$,
with only first time derivatives.
Dirac's constrained quantization
can be applied straightforwardly.
%with the momentum $p_{\mu(i)}$ conjugate to $x^{\mu(i)}$
%found to be
%\be
%p_{\mu(i)} = \sum_{k=0}^{N-i-1}\left(-\frac{d}{d\tau}\right)^k
%\frac{\del L}{\del x^{\mu(i+k+1)}}.
%\ee
%The Poisson bracket of
%any two $x^{\mu(n)}$'s vanishes
%if either both have $n < N$ or both have $n \geq N$.
%
%The Hamiltonian is
%\be
%H = \sum_{i=1}^{N-1} p_{\mu(i-1)}x^{\mu}_i
%+ p_{\mu(N-1)}\dot{x}^{\mu}_{N-1} - L,
%\ee
%where $\dot{x}^{\mu}_{N-1}$ is a function of
%$p_{(N-1)}$ and $x,x_1,\cdots,x_{N-1}$
%determined by
%\be
%p_{(N-1)} = \frac{\del L}{\del \dot{x}_{N-1}}.
%\ee
%The only dependence of $H$ on $p_i$
%for $0\leq i\leq N-2$ is linear,
%implying that $H$ is unbounded from below.

The equations of motion of the $x$'s are
differential equations of order $2N$,
which require $2N$ initial data to determine a solution.
One can think of the phase space as the space of
initial conditions given at $\tau = 0$.
Hence the phase space for each $x$ is $2N$ dimensional.

The fact that we have a larger phase space
for higher derivative theories has
significant implications.
Consider the worldline reparametrization symmetry (\ref{repara}).
Taking its $k$-th derivative and evaluating it at $\t = 0$,
we get
\be
\d x^{\mu(k)}(0) =
\sum_{m=0}^k C^k_m \eps^{(m)}(0) x^{\mu(k-m+1)}(0),
\;\; \mbox{where} \;\;
C^n_m = \frac{n!}{m!(n-m)!}.
\label{deltax}
\ee
Although these expressions follow directly from (\ref{repara}),
they are independent transformations on the phase space
for $k = 0, 1, 2, \cdots, N-1$,
since $x$, $\dot{x}$, etc. are independent variables.
Eq.(\ref{deltax}) defines $N$ independent transformations
with parameters $\eps(0), \dot{\eps}(0), \cdots, \eps^{(N-1)}(0)$.

Naively, even for $k \geq N$, eq.(\ref{deltax}) still
looks like a transformation on the phase space.
But it is inconsistent to treat them
as independent transformations
because they do not preserve the Poisson structure.
In other words,
it is impossible to find operators to generate these transformations.

The above can be derived more rigorously as follows.
We add (\ref{L1}) to the Lagrangian
and replace $x^{(n)}$'s by $x_n$'s.
The Lagrangian now has only first derivatives of $x$.
In order for $L'$ to be invariant
under the reparametrization,
we need
\be
\delta \lam_k = \frac{d}{d\tau}
\left[ \sum_{m=0}^{N-k-1} C^{k+m}_{m+1} \eps^{(m)} \lam_{k+m} \right],
%\label{lamt}
\quad
\delta x_k = \sum_{m=0}^k C^k_m \eps^{(m)} \dot{x}_{k-m}.
\label{xt}
\ee
The 2nd transformation law in (\ref{xt})
agrees with (\ref{deltax})
and can be conveniently summarized as
\be \label{xts}
\delta x(t+\s) = \eps(t+\s)\dot{x}(t+\s),
\ee
where we only need the first $N$ terms
in the Taylor expansion
\be
x(\t+\s) = \sum_{n=0}^{\infty} \frac{\s^n}{n!}\; x_n(\t), \quad
\eps(\t+\s) = \sum_{n=0}^{\infty} \frac{\s^n}{n!}\; \eps^{(n)}(\t).
\ee
Note that $\lam_n$ is identified with
$p_{n-1}$, the conjugate momentum of $x^{(n-1)}$,
in the constrained quantization.
So (\ref{xt}) also tells us how the momenta transforms.

The $N$ symmetry generators corresponding to $\eps^{(k)}$ are
\be \label{Hk}
H_k \equiv \sum_{m=k}^{N-1} C^{m}_k x^{\mu(m-k+1)} p_{\mu(m)}
\quad \mbox{for} \quad k = 0, \cdots, N-1.
\ee
We see that the reparametrization symmetry
is $N$ dimensional for a particle theory with
$N$-th time derivatives.

A redefinition of the $H_k$'s (\ref{Hk}) to
$L_k \equiv i(k+1)!H_{k+1}$ ($k = -1, 0, 1, \cdots, N-2$)
leads to
\be \label{Weyl}
[L_m, L_n] = (m-n) L_{m+n},
\quad m, n = -1, 0, 1, \cdots, N-2.
\ee
Note that this is a consistent truncation
of the classical Virasoro algebra.
This is one of the main results of this paper.

Roughly speaking, in the $N\rightarrow\infty$ limit,
we expect to obtain only ``half'' of the Virasoro algebra.
In order to have the full algebra,
one has to consider nonlocal theories
which are not defined as the large $N$ limit
of finite derivative theories.

\section{Conformal Symmetry for Nonlocal Particle}
\label{Conformal}

In this section we consider generic nonlocal particle theories.
%In this section we will take the limit
%of infinite derivatives
%\be \label{infN}
%N \rightarrow \infty.
%\ee
%In specific examples, one should be careful about
%how to take the limit when the infinite derivative theory
%is defined by a sequence of Lagrangians with
%finite order derivatives.
%Different limits may correspond to different theories,
%which may or may not be well defined.
%We take the viewpoint that these issues should be
%addressed in specific examples.
As various kinds of nonlocal theories
may differ significantly in nature,
in order for a generic discussion
without specifying details of the theory,
this section will be more heuristic and less rigorous than before.
We do not expect everything in this section
to hold for {\em all} nonlocal theories,
but we believe that our results will apply to
a wide class of examples.
In fact, it will be very interesting to have
{\em any} nonlocal particle theory for which
the reparametrization symmetry coincides with
the conformal symmetry.

To deal with Lagrangians with infinite derivatives,
an interesting proposal \cite{LV,Woodard,Gomis}
is to introduce an auxiliary coordinate $\s$ as follows.
For $x(\t)$ we introduce $X(\t, \s)$ and
impose the constraint
\be
\label{Xx}
\dot{X}(\t, \s) = X'(\t, \s),
\ee
where $X' \equiv \frac{d}{d\s}X$.
This guarantees that
$X$ can be identified with the original variable
as $X(\t, \s) = x(\t + \s)$.
Since the generic action (\ref{totalL}) also
involves higher derivatives of the tetrad $e(\t)$,
it should also be promoted to $E(\t, \s)$,
satisfying a constraint like (\ref{Xx}).
The covariant derivative $D$ (\ref{D})
will be replaced by
$D_{\s} = E^{-1}(\t, \s) \del_{\s}$.
when acting on $X$.
By replacing all $D$ by $D_{\s}$ and $x(\t)$ by $X(\t,\s)$ in $L_0$,
we obtain a new Lagrangian $L_1$
which has the same equation of motion
if the constraint (\ref{Xx}) is satisfied.

The constraint (\ref{Xx}) can be imposed by adding
\be
L' = \int d\s \; \lam(\s) \; ( X'(\s) - \dot{X}(\s) )
\ee
to $L_1$.
This can also be obtained as
an $N\rightarrow\infty$ limit of (\ref{L1}).
The new Lagrangian $L \equiv L_1 + L'$
does not have any higher time derivatives,
but only higher derivatives of the auxiliary coordinate $\s$.

The worldline reparametrization (\ref{reparatau})
of the original Lagrangian $L_0$ induces
the reparametrization of $\s$
\be \label{repara+}
\delta \s = - \eps(\t+\s),
\ee
leaving $\t$ invariant.
The derivative $D_{\s}$ and the scalar field $X$
should be invariant under (\ref{repara+}),
that is,
$f(\t, \s) = (f+\d f)(\t, \s+\d \s)$
for $f=D_{\s}$ and $X$.
It follows that
the transformation of $E(\t,\s)$ and $X(\t,\s)$ are given by
\bea
\d E(\t,\s) &=&  \del_{\s} \left( \eps(\t+\s) E(\t,\s) \right),
\label{Etrans} \\
\d X(\t,\s) &=& \eps(\t+\s) \del_{\s} X(\t,\s).
\label{Xtrans}
\eea
To have $L_1$ invariant under (\ref{repara+}),
we also need the measure of integration
$E(\t,\s)d\t\wedge d\s$ to be invariant,
which can be easily verified.

For $L'$ to be invariant, we need
\be \label{lamtrans}
\d \lam(\t,\s) = \del_{\s} \left( \eps(\t+\s) \lam(\t,\s) \right).
\ee
In addition, we need to impose certain boundary conditions
at the boundary values of $\s$ (denoted $\s_0$, $\s_1$) such that
\be
\left. \lam(\t,\s)(X' - \dot{X}) \right|_{\s_0}^{\s_1} = 0.
\ee
The boundary condition can be chosen to be
\be \label{BC}
\lam(\t,\s) = 0, \quad \mbox{for}
\quad \s = \s_0, \s_1, \;\; \forall \t.
\ee
This is also an appropriate condition to guarantee that
the equation of motion for the new Lagrangian $L$
is equivalent to the original equation of motion.
By varying $X$ and $\lam$ in the new Lagrangian $L$,
we get (\ref{Xx}) and
\be \label{eom2}
\dot{\lam}-\lam' = (\mbox{EOM}),
\ee
where $(\mbox{EOM})$ is the expression for
the equation of motion for the original Lagrangian $L_0$,
but with $x$ replaced by $X$, etc.
Using (\ref{Xx}) we can replace $X(\t,\s)$ by $x(\t+\s)$,
but (\ref{eom2}) does not guarantee that $(\mbox{EOM})=0$.
Instead, because $(\mbox{EOM})$ is a function of $(\t+\s)$ only,
(\ref{eom2}) implies that
\be
\lam(\t, \s) = (\t-\s) (\mbox{EOM}) + f(\t+\s)
\ee
for an arbitrary function $f$.
Applying the boundary condition (\ref{BC}) to it,
we find two identities valid for all $\t$.
They imply $\lam = 0$ and $(\mbox{EOM}) = 0$ as we hoped.

As the reparametrization symmetry (\ref{repara+})
is labelled by a one-variable function $\eps$ at $\t = 0$,
it is locally the same as the conformal symmetry in two dimensions.

So far we have restricted ourselves to
the classical theory.
In principle the quantum anomaly
(the central charge of the Virasoro algebra)
can appear in the large $N$ limit,
but we need further details of the theory
in order to treat it with some rigor.

\section{Remarks}
\label{Remarks}

{\noindent \em Connection to String Theory:}

Naively, higher derivative particle Lagrangians can be
related to the boundary string field theory \cite{Witten}
as the open string worldsheet Lagrangian
in the zero metric limit.
If this interpretation is acceptable,
the boundary string field theory can be interpreted
as the theory over ``the space of all particle theories''
instead of ``the space of all open string boundary theories''.
One can also view strings as a convenient technique to
summarize the infinite degrees of freedom of particles
with nonlocal interactions.
However,
it is hard to make this story rigorous due to
the technical problem that
higher derivative terms in the string worldsheet action
are nonrenormalizable.

Another possible connection with string theory
is to generalize the Seiberg-Witten (SW) limit \cite{SW}
for open strings ending on D-branes
in a constant $B$ field background.
Since the open string vertex operator
for the $U(1)$ field interaction
involves only the first time derivative,
only the zero mode of the string survives in the SW limit.
If other open string vertex operators
with higher derivatives are considered,
more degrees of freedom of the string will survive
the analogous SW limit, if it exists.

{\noindent \em Symmetry of Spacetime Fields}

In the infinite derivative limit,
we can generalize the story of general covariance
and $U(1)$ symmetry mentioned in the paragraph
below eq.(\ref{repara}) in sec.\ref{Review}.
The particle Lagrangian (\ref{totalL}) is invariant under
a simultaneous worldline field redefinition
\be \label{gt}
x \rightarrow x' = x'(x, \dot{x}, \ddot{x}, \cdots)
\ee
and a complicated, mixed transformation of the fields $A^{(P)}$.
%(The transformations generated by $Q_m$ in sec.\ref{Example}
% (\ref{Qf})
%is the residual symmetry of (\ref{gt})
%for the quadratic background.)
This defines a generalized notion of general covariance.
Similar tranformations for the string coordinates
has been used to construct string field theory \cite{Bardakci},
and it would be interesting to see if they have any connection
with the higher spin gauge theories \cite{higherspin}.

Generalization of the $U(1)$ symmetry is straightforward.
The Lagrangian is always defined only up to a total derivative
$ \frac{d}{d\t}\Lam(x, \dot{x}, \ddot{x}, \cdots) $,
which induces a gauge transformation of the $A^{P(n)}$'s.

{\noindent \em Stability Problem}

Higher derivative theories are known to have
various problems such as stability and causality.
Since the notion of causality may be changed
at the Planck scale where spacetime is fuzzy,
we will only comment on the problem of stability.
Theories with higher derivatives of finite order
suffer the Ostrogradskian instability
because the canonical Hamiltonian is always
unbounded from below \cite{Ostro}.
For low energy effective theories,
the perturbative formulation \cite{CHY} might be
a convenient way to avoid these problems.
On the other hand, stable nonlocal theories are known to exist.
At least those theories obtained from
integrating out certain physical fields
of a stable field theory should still be stable.

In this paper we studied worldline theories
with higher derivatives.
Although the Hamiltonian is constrained to vanish,
and thus is not unbounded from below,
an attempt to couple it to other physical systems
may still lead to instability.
Ideally, this work should have been focused on
nonlocal theories without instability.
We hope this paper may serve as part of the motivation
for the task of understanding how to construct
sensible nonlocal theories.

{\noindent \em Note Added in Revision}

After the first version of this paper appeared,
two earlier works \cite{Naka,Kato}
were brought to my attention.
These works considered nonlocal particle Lagrangians
closely related to the explicit example of this paper.
The work in Ref.\cite{Naka} did not mention
reparametrization symmetry,
and their model is not exactly the same as the string.
In the other work \cite{Kato},
a quantization different from
the canonical quantization of the original variables
was chosen such that it is equivalent to the open string.
It was also argued there
that the reparametrization symmetry
is equivalent to the conformal symmetry
for quadratic Lagrangians.
In this paper, we have provided a comparatively
more general and more explicit discussion
on the reparemetrization symmetry.

I was also informed that,
in a series of papers \cite{Segal},
generic particle Lagrangians
with only first derivatives are considered
as the starting point of formulating
the ``generalized equivalence principle''.
Our proposal for ``generalized general covariance''
described above is simply a generalization
of that to include higher derivatives.
This is a crucial difference when comparing
the corresponding higher spin gauge theory
with string theory.
Only totally symmetrized tensor fields
appear in the first deriviative theory of \cite{Segal},
but we know that in string theory there are
many more tensor fields.
Our model contains all the tensor field degrees of freedom
in open string theory.

\section*{Acknowledgment}

The author thanks
K. Bardakci, T.-C. Cheng, O. Ganor,
J. Gomis, K. Kamimura, H.-C. Kao,
M. Kato, J.-C. Lee, H. Ooguri,
S. Ramgoolam, K. Skenderis, W. Taylor
and M.-C. Yeh for helpful discussions,
and thanks the Aspen Center for Physics,
where part of this work was done.
This work is supported in part by
the National Science Council,
the Young Researcher Award of Academia Sinica,
the Center for Theoretical Physics
at National Taiwan University,
and the CosPA project of the Ministry of Education, Taiwan, R.O.C.

\end{document}